\documentstyle[12pt]{article}
\newcommand{\beq}{\begin{equation}}
\newcommand{\eeq}{\end{equation}}
\newcommand{\bra}{\begin{array}}
\newcommand{\era}{\end{array}}

\newcommand{\Om}{\Omega}

\begin{document}
\begin{center}
{\Large\bf Extended Weyl-Heisenberg algebra and Rubakov-Spiridonov
superalgebra: Anyonic realizations.}
\vskip1.4truecm

{\bf M. Daoud}\footnote{Permanent Address: Ibn Zohr, L.P.M.C.,
Departement of Physics, B.P. 28/S, Agadir, Morocco.} and {\bf J.
Douari}\footnote{douarij@ictp.trieste.it}
\vskip.2truecm
The Abdus Salam ICTP - strada costiera 11, 34100 Trieste, Italy.
\vskip1.4truecm

\end{center}
\vskip 2cm
\begin{abstract}
\hspace{.3cm}We give the realizations of the extended Weyl-Heisenberg
(WH) algebra and the Rubakov-Spiridonov (RS) superalgebra in terms of
anyons, characterized by the statistical parameter $\nu\in[0,1]$, on
two-dimensional lattice. The construction uses anyons defined from usual
fermionic oscillators (Lerda-Sciuto construction). The anyonic realization of
the superalgebra $sl(1/1)$ is also presented.
\end{abstract}
\newpage
\section{Introduction}
\hspace{.3cm}Anyons are particles with any statistics that interpolate
between bosons and fermions \cite{1,2,3}. They exist only in two dimension
because the configuration space of collection of identical particles has
some special topological properties allowing arbitrary statistics. On the
other hand supersymmetry (underlining  $Z_2$-graded superalgebras)
provided us with an elegant symmetry between fermions and bosons
\cite{4,5}. Due to his success, there have been many variances to
generalize its structures to incorporate other kinds of statistics. The first
attempt in this sens was proposed by Rubakov and Spiridonov by combining
bosonic and parafermionic degrees of freedom leading to the so-called
parasupersymmetric quantum mechanics of order 3 \cite{6}.The
3-fractional Rubakov-Spiridonov superalgebra has been extended to an
arbitrary order $k$ by Khare \cite{7} generalizing the $Z_2$ by a
$Z_k$-grading.The formalism of parasupersymmetry involves a bosonic   
degree of freedom (described by a complex variable) and a parafermionic
degree of freedom (described by a generalized Grassmman variable of
order $k$ \cite{8,9}). In other words, to pass from a $Z_2 $-graded theory to
a $Z_k$-graded one, we retain the bosonic variable and replace the  
fermionic variable by a parafermionic one. We note that the $k$-fermionic
variables \cite{10,11} have been also used to extend the
Rubakov-Spiridonov algebra \cite{12}. Fractional supersymmetry was also
developed without any explicit introduction of parafermionic or $k$-fermionic
degrees of  freedom. In this respect, fractional supersymmetry was worked by
Quesne and Vansteenkiste \cite{13} owing to the introduction of the extended
Weyl-Heisenberg algebra\cite{14} ( called also extended oscillator
 algebra).\\

The connection between anyons and fractional supersymmetry seems
apparently two very distinct subjects and have not been considered previously
in the literature. However, as we will see the statistical parameter of anyons
and the order of the fractional supersymmetry are deeply related.This
connection is based on the anyonization of the extended Weyl-Heisenberg
algebra. Then, considering the creation and annihilation anyonic operators
(called also anyonic oscillators), constructed by Lerda and Sciuto on a two
dimensional lattice \cite{15}, we disscus the anyonic realization of the
extended Weyl-Heisenberg algebra. The latter realization will be the
cornerstone to provide an anyonic realization of the fractional
Rubakov-Spiridonov superalgebra.\\

The present letter is organized as follows. First, we recall basis notions
connected with the Lerda-Sciuto anyonic oscillators on a two dimensional
Lattice. In section 3, we show the usefulness of the anyonic creation and
annihilation operators to provide a new realization of the extended
Weyl-Heisenberg algebra involving objects with fractional spin. Section 4 is
devoted to derivation of the fractional Rubakov-Spiridonov superalgebra
using the generators of the Weyl-Heisenberg algebra. In section 5, we show
that the anyons can be also used to realize the superalgebra $sl(1/1)$ 
which is undeformed in the quantum superalgebra context \cite{16,17,18} and not fractionnal
in the Rubakov-Spiridonov language.
\section{Basics tools: Lerda-Sciuto anyonic oscillators}
\hspace{.3cm}In this section, we recall necessary minimum of details
concerning anyonic oscillators  (see the reference \cite{15,16}) on $2d$ square
lattice $\Omega$ with spacing $a=1$. We give a two-component fermionic
spinor field by
\beq
S^- =\pmatrix{s_1 ^- (x)\cr s_2 ^- (x)\cr},
\eeq
and its conjugate hermitian by
\beq
S^+=(s_1 ^+ (x),s_2 ^+ (x)),
\eeq
such that the components of these fields satisfy the following standard
anticommutation relations
\beq
\bra{rcl}
\lbrace s_i ^- (x),s_j ^- (y)\rbrace &=&0\\
\lbrace s_i ^+ (x),s_j ^+ (y)\rbrace &=&0\\
\lbrace s_i ^- (x),s_j ^+ (y)\rbrace &=&\delta_{ij}\delta(x,y),
\era
\eeq
for $i,j \in\{1,2\}$ and $x,y\in\Omega$. Here, $\delta(x,y)$ is the
conventional lattice $\delta$-function: $\delta(x,y)=1$ if $x=y$ and vanishes
if $x\ne y$. We use the definition of the lattice angle functions   
$\Theta_{\pm\Gamma_{x}}(x,y)$ as recited in \cite{16}; where $\Gamma_{x}$
is the  curve associated to each site $x\in\Omega$ and the signs $\pm$
indicate the two kinds of rotation direction on $\Omega$. This definition
can be identified to that is discussed in \cite{15} by Lerda and Sciuto.\\

The expression of anyonic oscillators are given in term of fermionic spinors
and angle functions as follows
\beq
\bra{rcl}
a_i ^- (x_{\pm})&=&e^{i\nu\Delta_i (x_{\pm})}s_i ^- (x)\\
a_i ^+ (x_{\pm})&=&s_i ^+ (x)e^{-i\nu\Delta_i (x_{\pm})}.
\era
\eeq
The number $\nu$ appearing in this equation is usually called statistics
parameter. The elements $\Delta_i (x_{\pm})$ are given by
\beq
\Delta_i (x_{\pm})=\sum\limits_{y\in\Omega}s_i ^+ (x)\Theta_{\pm\Gamma_{x}}(x,y)s_i ^- (y),
\eeq
which satisfy the following commutation relations $$[\Delta_i (x_\pm),\ s_j ^-
(y)]=-\delta_{ij}\Theta_{\pm \Gamma_x}(x,y) s_i ^- (y)$$ $$[\Delta_i (x_\pm),\
s^+ _j (y)]=\delta_{ij}\Theta_{\pm \Gamma_x}(x,y)s^+ _- (y)$$ $$[\Delta_i
(x_\pm),\ \Delta_j (y_\pm)]=0$$ Now, we can show that the anyonic
oscillators satisfy the following algebraic relations
\beq
\bra{llllllllll}
\lbrack a_i ^- (x_{\pm}),a_i ^- (y_{\pm})\rbrack_{\Lambda^\mp}=0,& x>y\\
\lbrack a_i ^- (x_{\pm}),a_i ^+ (y_{\pm})\rbrack_{\Lambda^\pm}=0,& x>y\\
\lbrack a_i ^+ (x_{\pm}),a_i ^- (y_{\pm})\rbrack_{\Lambda^\pm}=0,& x>y\\
\lbrack a_i ^+ (x_{\pm}),a_i ^+ (y_{\pm})\rbrack_{\Lambda^\mp}=0,& x>y\\
\lbrack a_i ^- (x_{\pm}),a_i ^+ (x_{\pm})\rbrack=1,& \\
\lbrack a_i ^- (x_{\pm}),a_j ^+ (y_{\pm})\rbrack=0,& i\ne j\\
\lbrack a_i ^+ (x_{\pm}),a_j ^- (y_{\pm})\rbrack=0,& i\ne j\\
\lbrack a_i ^+ (x_{\pm}),a_j ^+ (y_{\pm})\rbrack=0,& i\ne j\\
\lbrack a_i ^{\pm} (x_{-}),a_j ^{\pm} (y_{+})\rbrack=0,& \forall i,j\\
\lbrack a_i ^- (x_{-}),a_j ^+
(y_{+})\rbrack=\delta_{ij}\delta(x,y)\Lambda^{+[\sum\limits_{z<x}-\sum\limits_{z>x}]s_j
^+ (z)s_j ^- (z)},&\\
\lbrack a_i ^- (x_{-}),a_j ^+
(y_{+})\rbrack=\delta_{ij}\delta(x,y)\Lambda^{-[\sum\limits_{z<x}-\sum\limits_{z>x}]s_j
^+ (z)s_j ^- (z)}.&
\era
\eeq
where $\Lambda^\pm =e^{\pm i\nu\pi}$, $[X,Y]_\Lambda = XY+\Lambda YX$ and
$$x>y\Leftrightarrow
\left\{\begin {array}{l}
x_+>y_+\Leftrightarrow\left\{\begin{array}{l}
x_2>y_2\\[2mm]
x_1>y_1, x_2 =x_1\end{array}\right.\\[3mm]
x_-<y_-\Leftrightarrow
\left\{\begin{array}{l}
x_2<y_2\\[2mm]
x_1<y_1, x_1 =x_2\end{array}\right.\end{array}\right. $$
One obtains also
\beq
(a_i ^\pm (x_\pm))^2 =0,
\eeq
which is known as the hard core condition.\\

We would like to stress that despite the many formal analogies, the anyonic
oscillators does  not have anything to do with the $k$-fermions,
$q$-deformed bosons with $q=e^{2{\pi}i/k}$, for  several reasons: $(i)$ the
$k$-fermions can be defined in any dimensions whereas the anyons are 
strictly two-dimensional objects, $(ii)$ the anyons are non-local contrarly to
the $k$-fermions.  The latter objects constitute a mathematical tool,
introduced in the context of quantum algebras,  which was used to go
beyond the conventional statistics in any dimension and taking into account
some  perturbation (deformation) responsible of small deviations from the FD
and BE usual statistics \cite{19,20}.
\section{Anyonic realization of the extended Weyl-Heisenberg algebra}
\hspace{.3cm}For fixed $\nu\in[0,1]$, the extended Weyl-Heisenberg algebra
is defined as an algebra generated by the operators $a_+$, $a_-
=(a_+)^+$, $N$ and $K=(K^+ )^{-1}$ satisfying the following  relations
\cite{12,13}
\beq
\bra{lll}
\lbrack a_- ,a_+ \rbrack =\sum\limits_{s=0}^{\frac{2}{\nu}-1}f_s P_s,& \lbrack
N
,a_{\pm}\rbrack =\pm a^{\pm} \\
\\
Ka_+ = e^{i\pi\nu}a_+ K, & Ka_- = e^{-i\pi\nu}a_- K\\
\\
\lbrack K ,N \rbrack = 0,& K^{\frac{2}{\nu}}=1.
\era
\eeq
Here, $f_s$ are some real parameters and the admissible values of the
statistical parameter is restricted by the condition $\frac{2}{\nu}\in\{2,3,4...\}$.
The operators $P_s$ are polynomials in the grading operator $K$ defined by
\beq
P_s = \frac{\nu}{2}\sum\limits_{t=0}^{\frac{2}{\nu}-1}e^{-i\pi\nu st}K^t ,
\eeq
for $s=0,1,2,...,\frac{2}{\nu}-1$. One can see that the $P_s$'s operators
satisfy the folowing relations
\beq
\bra{rcl}
\sum\limits_{t=0}^{\frac{2}{\nu}-1}P_s =1,& P_s P_t = \delta (s,t) P_s 
\era
\eeq
where $\delta$ is the Kronecker symbol. Furthermore, these operators satisfy
\beq
\bra{rcl}
P_s a^+ = a^+ P_{s-1},& a^- P_s =P_{s-1}a^- .
\era
\eeq
Note that the equation (9) can be conversed in the form
\beq
K^t = \sum\limits_{s=0}^{\frac{2}{\nu}-1}e^{i\pi\nu st}P_s ,
\eeq
with $t=0,1,...,\frac{2}{\nu}-1$. It is clear that the commutation relation
between $a_-$ and $a_+$ (equations (8)) can be written as
\beq
\lbrack a_- ,a_+ \rbrack =\sum\limits_{s=0}^{\frac{2}{\nu}-1}c_s K^s 
\eeq
where the $c_s$ are related to parameters $f_t$ by
\beq
f_t =\sum\limits_{s=0}^{\frac{2}{\nu}-1}e^{i\pi\nu st}c_s 
\eeq
or conversely by
\beq
c_s = \frac{\nu}{2}\sum\limits_{t=0}^{\frac{2}{\nu}-1}f_t e^{-i\pi\nu st}.
\eeq
To show that the extended Weyl-Heisenberg algebra can be realized by
means of anyonic oscillators of  statistics $\nu$ in a quite direct though
non-trivial way, we start by introducing the local operators
\beq
\bra{rcl}
a_+ (x)&=&a_1 ^+ (x_{+})a_2 ^- (x_{+})\\
a_- (x)&=&a_2 ^+ (x_{-})a_1 ^- (x_{-})
\era
\eeq
and
\beq
N(x)=a_1 ^+ (x_{+})a_1 ^- (x_{+})-a_2 ^+ (x_{+})a_2 ^- (x_{+}).
\eeq
It is not difficult to verify, from the commutation relations (6), that 
\beq
\bra{rcl}
a_+ (y)a_+ (x)&=&e^{i2\pi\nu}a_+ (x)a_+ (y)\\ \\
a_- (y)a_- (x)&=&e^{-i2\pi\nu}a_- (x)a_- (y)
\era
\eeq
for $x>y$. The latter equation show that the local operators defined by (16)
have braiding properties and one have to specify the ordering of the points
$x$ and $y$ if we change the braiding orientation. In the same way, with a
straighfoward application of equations (6), we get
\beq
\bra{ll}
\lbrack N(x) ,a_{\pm}(y) \rbrack =\pm a_{\pm}(x)\delta(x,y),&\\ \\
\lbrack a_+ (x) ,a_{-}(y) \rbrack =0,& x\ne y
\era
\eeq
The commutation relation of $a_+(x)$ and $a_-(x)$ (i. e. at the same point) is
slightly more complicated. However, by an adequate using of the commutation
relations of the anyonic oscillators, it is not hard to show that 
\beq
\lbrack a_+ (x) ,a_{-}(x) \rbrack =\prod\limits_{y<x}e^{-i\nu\pi
N(y)}N(x)\prod\limits_{z>x}e^{-i\nu\pi N(z)}.
\eeq Now, we introduce the global creation, annihilation and number
anyonic operators of spin $s=\frac{\nu}{2}$ on the lattice in terms of the local
anyonic ones. They are defined by
\beq
\bra{rcl}
a_+ &=&\sum\limits_{x\in\Om}a_1 ^+ (x_{+})a_2 ^- (x_{+})\\
a_- &=&\sum\limits_{x\in\Om}a_2 ^+ (x_{-})a_1 ^- (x_{-}),
\era
\eeq
and
\beq
N= \sum\limits_{x\in\Om}(a_1 ^+ (x_{\pm})a_1 ^- (x_{\pm})-a_2 ^+
(x_{\pm})a_2 ^- (x_{\pm}))
\eeq
The grading operator (known also as Klein operator) is defined by
\beq
K=e^{[i\pi\nu \sum\limits_{x\in\Om}(a_1 ^+ (x_{\pm})a_1 ^- (x_{\pm})-a_2 ^+
(x_{\pm})a_2 ^- (x_{\pm}))]}
\eeq

It is clear that the operator $K$ satisfy
\beq
K^{\frac{2}{\nu}}=1
\eeq
 and we have also
\beq
(a_+ )^+ =a_- .
\eeq
The operators $N$ and $K$ are commutating.  Furthermore using the
structure relations of the anyonic  oscillators on the lattice $\Omega$, we
have the following commutation relations
\beq
\bra{ll}
\lbrack a_+ ,a_- \rbrack
=\sum\limits_{x\in\Om}(\prod\limits_{y<x}e^{[-i\pi\nu
N(y)]})N(x)(\prod\limits_{z>x}e^{[-i\pi\nu N(z)]}),& \\ \\
\lbrack N,a_{\pm} \rbrack =\pm a_{\pm},& Ka_{\pm}=e^{\pm i\nu\pi}a_{\pm}K. 
\era
\eeq
To write the commutation relation involving $a_-$ and $a_+$ (Eqs (26)), in a form similar to the one given in
equations (8),  we show that
\beq
\bra{rcl}
\sum\limits_{x\in\Om}(\prod\limits_{y<x}e^{[-i\pi\nu
N(y)]})N(x)(\prod\limits_{z>x}e^{[-i\pi\nu N(z)]})&=&-\sum\limits_{t=0}^{\frac{2}{\nu}-1}f_t
P_t \\
&=&-\sum\limits_{t=0}^{\frac{2}{\nu}-1}c_s K^s
\era
\eeq
where the parameters $f_t$ and $c_t$ are given by
\beq
f_t = - \frac{\sin(2\pi\nu t)}{\sin(\pi\nu )}
\eeq
and
\beq
c_s = \left\{\bra{lll}
0,& s\ne 2,\frac{2}{\nu}-2\\
\frac{1}{2\sin(\pi\nu)},& s=2\\
-\frac{1}{2\sin(\pi\nu)},& s=\frac{2}{\nu}-2
\era\right.
\eeq

To obtain this result, we followed a similar method that one established by
Lerda and Sciuto in \cite{15}. The first step of this method is based on the
fact that the local operator $N(x)$ admits only the eigenvalues $0$ and $+1$
for any $x\in\Om$ according to the Pauli exclusion principle for anyon
operators (hard core condition). Then, one have the identity
\beq
N(x) = \frac{\sin(\pi\nu N(x))}{\sin(\pi\nu )}
\eeq
for any site $x\in\Omega$. Reducing the lattice $\Omega$ to only one site,
$x_0 $ for instance, we have $N=N(x_0)$ and thus the equality (27) hold
thanks to equation (30).\\

Assuming that the equation (27) is valid for a lattice of $n$ sites $\{x_i ,
i=1,...n\}$, we add an extra  point $x_{n+1}$. For the lattice with $(n+1)$
sites, the equation (27) becomes
\beq
\sum\limits_{i=1}^{n}(\prod\limits_{j<i}e^{-i\pi\nu N(x_j )}N(x_i
)\prod\limits_{k>i}e^{-i\pi\nu N(x_k )})e^{N(x_{n+1})}+e^{i\pi\nu
N(x_{n+1})}N(x_{n+1})=\frac{\sin(\pi\nu N)}{\sin(\pi\nu)}
\eeq
where $N=\sum\limits_{i=1}^{n+1}N(x_i )$. The last equation is easily
designed using the equation (26) and (30). In view of the relations (24), (26) and
(27), we conclude that the operators $a_+$, $a_-$, $N$ and $K$ close the extended 
Weyl-Heisenberg algeba.\\

The operators $a_- ^{\frac{2}{\nu}}$, $a_+ ^{\frac{2}{\nu}}$ and $K
^{\frac{2}{\nu}}$ belong the centre of the extended Weyl-Heisenberg algebra.
It is straightforward to prove also that the operator
\beq
C=a_- a_+ +\frac{\sin(\pi\nu(2N+1))}{2\sin^2 (\pi\nu)}
\eeq
is an invariant of the WH algebra. Then, the extended WH algebra admits
finite-dimensional representations of dimension $\frac{2}{\nu}$ such that
\beq
a_- ^{\frac{2}{\nu}} =\alpha\bf{1}\rm_{\frac{2}{\nu}\times\frac{2}{\nu}}
\eeq
and
\beq
a_+ ^{\frac{2}{\nu}} =\beta \bf{1}\rm_{\frac{2}{\nu}\times \frac{2}{\nu}}
\eeq
where $(\alpha,\beta)\in{\bf C^2}$. Three types of representation can be
considered: $(i)$ $\alpha=\beta=0$ (nilpotent representation), $(ii)$
$\alpha=\beta=1$ (cyclic or periodic representations) and $(iii)$ $\alpha=0$
and $\beta\ne 0$ or $\alpha\ne 0$ and $\beta=0$ (semi-periodic
representations). In a representation of type $(i)$, the creation $a_+$ and
annihilation $a_-$ seems to be similar to ones of
($k=\frac{2}{\nu}$)-fermions \cite{11,21,22} which satisfies the generalized
exclusion principle according to which no more than $(k-1)$ particles
can live in the same quantum state \cite{20}.
\section{Fractional RS superalgebra through fermionic anyons}

In view of the fact that supersymmetry (Models and symmetries with $Z_2$
-grading) provide symmetry between bosons and fermions, it is natural to
enquire if one could generalize these structures by including the exotic
statistics. More than ten years back \cite{6}, Rubakov and Spiridonov
discussed such generalizations. In particular, they constructed a
superalgebra defined by the following structure relations
\beq
\bra{lll}
(E_- )^3 =0\\ \\
\lbrack E_- , H\rbrack =0\\ \\
(E_- )^2 E_+ +E_- E_+ E_- +E_+ (E_- )^2 =2E_- H
\era
\eeq
and the hermitian conjugated relations. The RS superalgebra generated by
$\left\{E_+ ,E_- ,H\right\}$ was realized as $Z_3 -$graded symmetry
between one boson and one parafermion of order 2. The generalization of
RS superalgebra (37) for an arbitrary order $k$ is given by \cite{7}.
\beq
\bra{lll}
(E_- )^k =0\\ \\
\lbrack E_- , H\rbrack =0\\ \\
(E_- )^{k-1} E_+ +(E_-)^{k-2} E_+ E_- +E_+ (E_- )^{k-1} =(k-1)(E_- )^{k-2} H
\era
\eeq
together with their hermitian conjugates. The realization of the RS
superalgebra (36) involve one boson and $(k-1)$ parafermions. Note that for
$k=2$, the relations (36) reduces to ones defining the superalgebra $sl(1/1)$.
Then the RS superalgebra seems to be a $k$-fractional extension of the
superalgebra $sl(1/1)$.\\ 

In this part of our work, we will discuss how using the generators of extended
Weyl-Heisenberg algebra (remember that the operators of this algebra are
defined in terms of anyonic oscillators living on two-dimensional lattice) one
can give an anyonic realization of the so-called Rubakov-Spiridonov
superalgebra. Indeed, By means of the operators $P_i$'s (polynomials in
the grading operator $K$) and the creation and annihilation operators
$a_+$ and $a_-$ (that are defined by coupling two 2d-lattice anyons), we
are thus in position to define the operators
\beq
\bra{rcl}
E_- =a_- (1-P_{k-1}),& E_+ =a_+ (1-P_{0})
\era
\eeq
among $k$ possible definitions. The $(k-1)$ other definitions can be
obtained by a simple permutation of indices $0,1,2,3...,k-1.$ Note that the
operators $E_+$ and $E_-$ are defined such that satisfy also
\beq
\bra{rcl}
E_+  =(E_-)^+
\era
\eeq

As a first result, one can show, using the structure relations of the extended
Weyl-Heisenberg algebra introduced in the previous section, that
\beq
\bra{rcl}
(E_- )^m &=&(a_- )^m [1-P_{k-m}+P_{k-m+1}+...+P_{k-1}]\\
(E_+ )^m &=&(a_+ )^m [1-P_{0}+P_{1}+...+P_{m-1}]
\era
\eeq
for $m=1,2,...,k-1$. For the particular case $m=k$, we have
\beq
\bra{rcl}
(E_- )^k =0 ,& (E_+ )^k =0.
\era
\eeq
Let us note that for $k=2$, we have the nilpotency condition of the $sl(1/1)$
generators $E_- $ and $E_+ $. We continue with the construction of the
fractional Rubakov-Spiridonov superalgebra. So, one can obtain  also, by
some manipulations more or less complicated, the following identity
\beq
(E_- )^m E_+ =(a_- )^m a_+ (P_{1}+P_{2}+...+P_{k-m})
\eeq
which leads to  the relations
\beq
\bra{rcl}
(E_- )^m E_+ (E_- )^l =(a_- )^m a_+ (a_- )^l ,& m+l=k-1,
\era
\eeq
with $m\ne 0$; $l\ne k-1$ and  $m\ne k-1$; $l\ne 0$,
\beq
E_+ (E_- )^{k-1} =a_+ (a_- )^{k-1}P_0
\eeq
\beq
(E_- )^{k-1}E_+  =(a_- )^{k-1}a_+ P_1 .
\eeq

The equations (42),(43) and (44) are the basic ones to get the multilinear
relation which should be satisfied  by the the generators $E_-$ and $E_+$ 
of the Rubakov-Spiridonov superalgebra. Indeed, introducing an even 
operator $H$, we show that
\beq
\bra{rcl}
\sum\limits_{i=0}^{k-1}(E_- )^{k-1-i} E_+ (E_- )^i =(k-1)(E_- )^{k-2}H,
\era
\eeq
where $H$ is defined by
\beq
H=\sum\limits_{s=0}^{k-1}H_{k-s}P_s
\eeq
where
\beq
H_k =a_+ a_- -\frac{1}{k-1}\sum\limits_{s=2}^{k-1}\sum\limits_{t=1}^{s-1}h(N-t)
\eeq
and
\beq
\bra{rl}
H_{k-s}=H_k
+\frac{1}{k-1}\sum\limits_{t=0}^{k-2}\sum\limits_{i=1}^{s}h(N+i-1-t),& s\ne 0.
\era
\eeq
The function $h$ is given by
\beq
h(N-l)=\frac{\sin(\nu\pi (N-l))}{\sin(\nu\pi)}
\eeq
The operator $H$ commute with the generators $E_+$ and $E_-$
\beq
\lbrack H,E_{\pm}\rbrack =0.
\eeq
The nilpotency relations (40) toghether with the multilinear relation (45) and
the commutation relation (50) close the fractional Rubakov-Spiridonov
superalgebra. Let us note that this  superalgebra is different from one
obtained in the context of quantum superalgebra which is characterized by
the following structure relations \cite{17,18}
\beq
\bra{lll}
\lbrace E_+ ,E_- \rbrace =\frac{q^H -q^{-H}}{q-q^{-1}},&\\
\lbrack E_{\pm},H\rbrack =0,&\\
(E_+ )^2 =0,& (E_- )^2 =0.
\era
\eeq
Two limiting cases $\nu=1$ and $\nu=0$ are interesting. In the case where
$\nu=1$, it is easy to see the Rubakov-Spiridonov superalgebra gives the
well known superalgebra $sl(1/1)$. In the case $\nu =0$, the operator $K$ is
equal to unit operator and the projection operators $P_i $'s vanishes. Then,
we have
\beq
\bra{rcl}
E_+ =a_+ ,& E_- =a_- 
\era
\eeq
where the operator $a_+ $ and $a_- $ are defined in terms of two bosonic
oscillators  (because in the limiting case $\nu =0$, the anyons becomes
bosons). As consequence, $E_+$, $E_-$ and $H$  generate the $sl(2)$
algebra. Then, in the limit $\nu=1$, the RS superalgebra reduces to $Z_2
$-graded   $sl(1/1)$ superalgebra and for $\nu=0$, we have $sl(2)$ (no
graduation).\\

We note finally that the structure relations defining the RS superalgebra are
analogous to those defining  the so-called fractional supersymmetric
quantum mechanics \cite{6,7}. So, one can ask about the physical meaning
of the even operator $H$ and if one can interpret it as an hamiltonian
describing a system of two anyons constrained to evolve in a two
dimensional lattice. We believe that this point merit more investigation and
can help us to learn more about the physics in low dimensions.
\section{Anyonic realization of the  $sl(1/1)$ superalgebra}
In this last section, we would draw the attention that the superalgebra
$sl(1/1)$ (undeformed in the quantum superalgebra context and no
fractional in the Rubakov-Spiridonov sens) can be realized using two
anyons living on two dimentional lattice. In this order define the operators
\beq
\bra{rcl}
q_{i,+}&=&a_+ P_{k-i}\\ \\
 q_{i,-}&=&P_{k-i}a_-
\era
\eeq
and
\beq
h_i =a_- a_+ P_{k-i}+a_+ a_- P_{k-i+1}
\eeq
(for $i=2,3,...,k$) where the operators $a_+$ and $a_-$ are defined by (21)
in terms of the Lerda-Sciuto anyonic oscillators. A direct computation show
that operators $q_{i,+}$, $q_{i,-}$ and $h_i$ satisfy the structure relations
\beq
\bra{ll}
\lbrack q_{i,\pm}, h_i \rbrack =0,& \lbrace q_{i,-}, q_{i,+}\rbrace =h_i \\ \\
(q_{i,+})^2 =0,& (q_{i,-})^2 =0
\era
\eeq
that are ones defining the $sl(1/1)$ superalgebra. This result show how one
can realize the (undeformed) superalgebra $sl(1/1)$ using the anyonic
oscillators. Furthermore, our realization is new in the sens that not has been
considered previously in the literature and should not be confused with
Frappat et al \cite{18} (see also the references therein) realizations which
concerns  the anyonic representations of the quantum (deformed)
superalgebras. One can think that the realization obtained here of $sl(1/1)$
superalgebra, which is essentially a coupling the 2d lattice anyons, suggests 
some probable manifestation of the supersymmetry in low dimensional
physics.

It is important to note that the odd operators $E_+$ and $E_-$ (see Eqs (37))
can be expressed as sums of the odd operators $q_{i,+}$ and $q_{i,-}$,
generating (with $h_i$) the superalgebra $sl(1/1)$, as follows
\beq
\bra{rl}
E_+ =\sum\limits_{i=2}^{k}q_{i,+},& E_- =\sum\limits_{i=2}^{k}q_{i,-}
\era
\eeq
which are Ladder operators of the fractional Rubakov-Spiridonov
superalgebra (see the previous section).

\section{Conclusion}
\hspace{.3cm}
In the present letter, we introduced the anyonic realization of the extended
Weyl-Heisenberg algebra by means of Lerda-Sciuto anyonic oscillators of
statistics $\nu$ on two dimensional lattice. We discussed how the
$Z_\nu$-grading of this algebra leads to the realization of the fractional
Rubakov-Spiridonov superalgebra. We then proved the connection between
the fractional spin of anyons and the order of the RS superalgebra. We 
showed also that the superalgebra $sl(1/1)$ can be constructed following an
anyonic scheme.

It is clear that remain many open problems for future study. One of them
would be a better understanding of the physical meaning of the operator
H(Eq(46)) and if can it be related to the hamiltonian describing planar
systems evolving under topological Chern-Simon's interaction. Another
would concerns the anyonic realizations of the fractional analogous of the
other superalgebras (undeformed in the quantum groups language) than
$sl(1/1)$.These problems are under study and we hope to report on them in
the near future.
\section*{Acknowledgments}
The authors would like to thank the Abdus Salam International Center for
Theoretical Physics (ICTP), Trieste, for its warm hospitality. The author M. D would
like to express his sincere thanks to Prof. Yu Lu (head of Condensed Matter
in ICTP) for his kind invitation. The author J. D would acknowledges the Max-Planck-Institut f\"ur Physik
Komplexer Systeme for link hospitality during the stage in which one part of
this paper was done, and She particularly would like to thank the
associateship Sckeme in the Abdus Salam ICTP for its facilities.

\end{document}